\documentclass[aps,prb,reprint, superscriptaddress,showpacs]{revtex4-1}

\usepackage{graphicx}
\usepackage{dcolumn}
\usepackage{bm}

\begin{document}

\preprint{APS/123-QED}

\title{Finite shot noise and electron heating at quantized conductance in high-mobility quantum point contacts}

\author{Tatsuya Muro}
\affiliation{Graduate School of Science, Osaka University, Toyonaka, Osaka 560-0043, Japan}

\author{Yoshitaka Nishihara}
\affiliation{Graduate School of Science, Osaka University, Toyonaka, Osaka 560-0043, Japan}
\affiliation{Institute for Chemical Research, Kyoto University, Uji, Kyoto 611-0011, Japan}

\author{Shota Norimoto}
\affiliation{Graduate School of Science, Osaka University, Toyonaka, Osaka 560-0043, Japan}

\author{Meydi Ferrier}
\affiliation{Graduate School of Science, Osaka University, Toyonaka, Osaka 560-0043, Japan}
\affiliation{Laboratoire de Physique des Solides, CNRS, Univ. Paris-Sud, Universit\u{e} Paris Saclay, 91405, Orsay Cedex, France}

\author{Tomonori Arakawa}
\affiliation{Graduate School of Science, Osaka University, Toyonaka, Osaka 560-0043, Japan}

\author{Kensuke Kobayashi}\email{kensuke@phys.sci.osaka-u.ac.jp} 
\affiliation{Graduate School of Science, Osaka University, Toyonaka, Osaka 560-0043, Japan}

\author{Thomas Ihn}
\affiliation{Solid State Physics Laboratory, ETH Zurich, 8093 Zurich, Switzerland}

\author{Clemens R{\"o}ssler}
\affiliation{Solid State Physics Laboratory, ETH Zurich, 8093 Zurich, Switzerland}

\author{Klaus Ensslin}
\affiliation{Solid State Physics Laboratory, ETH Zurich, 8093 Zurich, Switzerland}

\author{Christian Reichl}
\affiliation{Solid State Physics Laboratory, ETH Zurich, 8093 Zurich, Switzerland}

\author{Werner Wegscheider}
\affiliation{Solid State Physics Laboratory, ETH Zurich, 8093 Zurich, Switzerland}

\date{published 9 May 2016}

\begin{abstract}
We report a precise experimental study on the shot noise of a quantum point contact (QPC) fabricated in a GaAs/AlGaAs based high-mobility two-dimensional electron gas (2DEG). xThe combination of unprecedented cleanliness and very high measurement accuracy has enabled us to discuss the Fano factor to characterize the shot noise with a precision of 1 \%. We observed that the shot noise at zero magnetic field exhibits a slight enhancement exceeding the single particle theoretical prediction, and that it gradually decreases as a perpendicular magnetic field is applied. We also confirmed that this additional noise completely vanishes in the quantum Hall regime. These phenomena can be explained by the electron heating effect near the QPC, which is suppressed with increasing magnetic field.
\end{abstract}

\maketitle

\section{\label{sec:level1}Introduction}
Quantum point contact (QPC) is a narrow constriction of the order of the Fermi wave length of electrons in a two-dimensional electron gas (2DEG) system\cite{vanwees,wharam}. Owing to its characteristic property that the conductance is quantized in units of $2e^2/h \sim (12.9\ \rm{k}\Omega)^{-1}$ due to the ballistic conduction and the wave nature of electrons,  QPC has attracted great attention as an ideal realization of Landauer picture of mesoscopic transport. In spite of the long research history, it still remains a considerable interest in terms of  many-body effect such as the 0.7 anomaly~\cite{thomas1,thomas2,thomas3,nuttinck,reilly,crook,yoon,chung,kritensen,cronenwett,lesovik}. Thus, to precisely clarify the mesoscopic transport in QPC from various points of view is still of scientific significance.

Shot noise, which is the Fourier transform of the current-current correlation function in the non-equilibrium regime, has been a useful probe to investigate the transport properties of QPC. Theoretically, this topic has been treated since early 1990's and participated to establish mesoscopic physics~\cite{lesovik,buttiker1,buttiker2,landauer,buttiker3,blanter}. The shot noise, which reflects the granularity of electrons, can be written $S_{shot}=2e\langle I\rangle F$, where $F$ is the Fano factor which characterizes the electron partition process. The non-interacting scattering theory~\cite{blanter} predicts $F=\sum_n T_n(1-T_n)/\sum _n T_n$ ($T_n$ is the transmission of $n$-th channel) and, as the QPC conductance is given by $G=2e^2/h\sum _nT_n$, we expect that the shot noise vanishes and Fano factor becomes zero for every integer multiple of $2e^2/h$. Since this theory is based on the single-particle picture, additional information such as the many-body interaction may be deduced from the precise investigation of the Fano factor combined with the conductance, as we see in the shot noise study in the Kondo regime~\cite{PhysRevB.77.241303,DelattreNatPhys2009,PhysRevLett.106.176601,FerrierNatPhys2016}.  Actually, the shot noise of QPC was addressed by several experimental groups; the Pauli exclusion principle was first studied, and then the quantum interference~\cite{liu}, the 0.7 anomaly\cite{roche,dicarlo,nakamura}, and the spin polarization\cite{dicarlo,kohda}.

\begin{figure}[b]
\includegraphics[width=.99\linewidth]{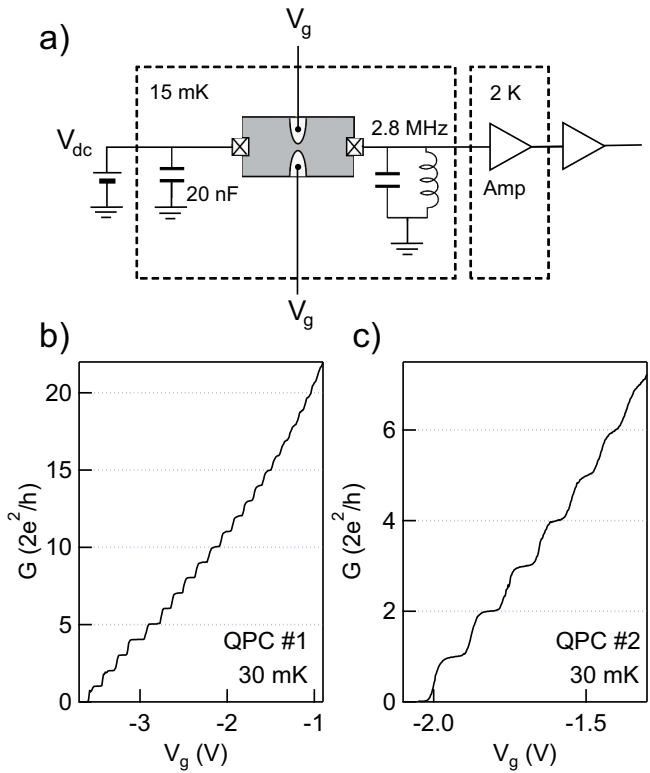}
\caption{\label{mesure}(a) Schematic diagram of the current noise measurement setup. The QPC is formed by negative voltage $V_g$ applied on two opposing electrostatic gates. (b) and (c)  Conductance of QPC 1 and 2 measured at 30 mK as a function of gate voltage. 
}
\end{figure}

In spite of various experimental attempts on the shot noise in QPC, many research groups have been repeatedly reporting an interesting observation, that is, an enhanced QPC shot noise exceeding theoretical value\cite{kumar,reznikov,roche,hashisaka,nakamura,nishihara,kohda}. This deviation was attributed to several reasons: the $1/f$ noise, the channel mixing, and the electron heating as the most likely one\cite{nishihara}. Previous studies on QPC addressed the problem by assuming a model\cite{kumar}, which phenomenologically takes energy relaxation around QPC into account. Although it is experimentally known that the enhanced Fano factor can be suppressed by applying a weak perpendicular magnetic field to 2DEG\cite{dicarlo,kumar}, there are no systematic studies on how magnetic field affects this phenomenon.

In this paper, we report an experimental study on a very precise shot noise measurement in ultra-high-mobility QPCs. The combination of an unprecedented clean QPC and an accurate noise measurement setup has enabled us to obtain the Fano factor with the precision of 1 \%. By systematically investigating the magnetic field and temperature dependence of the Fano factor at the conductance plateaus, we are able to discuss the energy dissipation process around QPC. Then we show that the electron heating effect is suppressed either by perpendicular magnetic field or by increasing the conductance of QPC. These observations are explained by a model with a single phenomenological parameter ($G_m$) which characterizes the heat transport. It is also confirmed that a perfectly noiseless transport is realized in the quantum Hall regime. Our experiments imply that the heat dissipation in a 2DEG system cannot be neglected even in ultra-clean conductors and that the electron-electron scattering or energy-loss mechanism can intrinsically affect transport phenomena in low-dimensional electron systems.

\section{Experiments}

\subsection{Conductance measurement}

We investigated two QPC devices (QPC 1 and QPC 2) fabricated on the same GaAs/AlGaAs heterostructure wafer with a 2DEG located 160 nm beneath the surface\cite{rossler,rossler2}. Each QPC is independently defined by applying voltage $V_g$ to each gate electrode. The distance between the gates to define the QPC is 500 nm and 200 nm for QPC 1 and 2, respectively~\cite{rossler}. The mobility and electron density of the 2DEG are $\mu=1000$-$2000\ \rm{m^2/Vs}$ and $n_s=3.5\times10^{15}\ /\rm{m^2}$, respectively. The Fermi wavelength of the 2DEG is 42 nm~\cite{rossler}. After bias-cooling with $V_g$=+1 V, the experiments were carried out in a dilution refrigerator whose base temperature is 15 mK. The differential conductance $G=dI/dV_{\mbox{sd}}$ is measured by standard lock-in technique as a function of the source-drain bias voltage ($V_{\mbox{sd}}$). The lead and the contact resistances in series with the sample were deduced by measuring the conductance with no gate voltage applied at every applied magnetic field and was subtracted accordingly.

Figure 1 shows the differential conductance of QPC 1 and QPC 2 at 30 mK as a function of $V_g$. The conductance steps up to the 20th (8th) are well resolved in QPC 1 (QPC 2). The conductance at the $n$-th plateau precisely agrees with $2e^2/h \times n\ (n=1, 2, 3,...)$ within the precision of $\pm 2\  \%$. This validates the above treatment for the lead and contact resistances. Based on this observation, we safely eliminate the possibility of channel mixing since it would result in non-integer conductance step. We also checked the temperature dependence of the conductance from $T$=15 mK to 4.2 K and found that the step structure does not depend on temperature below 550 mK. We observed an indication of the shoulder structure at $0.7\times 2e^2/h$ (0.7 anomaly) at 2 K and it becomes prominent above 4 K. However, the 0.7 anomaly is beyond the scope of the present work.

\subsection{Current noise measurement}

In addition to the conductance measurement, we performed current noise measurements as follows\cite{arakawa,nishihara}. The voltage fluctuation at 2.8 MHz defined by the resonant circuit is extracted as an output signal of the homemade cryogenic amplifier\cite{arakawa}. The time-domain voltage noise signal is then captured by a digitizer and converted by fast-Fourier-transform (FFT). The spectral density of current fluctuation $S_I$ is obtained by fitting the resonance peak $P_0$,
\begin{equation}
P_0=A\left[X_V+\left(\frac{Z+R}{ZR}\right)^2(S_I+X_I)\right]
\end{equation}
where $R$ is the measured differential resistance, $A$ is the gain of the cold and room temperature amplifiers, $Z$ is the impedance of the measurement circuit, and $X_{I} (X_{V})$ is  current (voltage) noise due to the amplifier, respectively. The typical values are $Z=3.2\times10^4, A=1.4\times10^6, X_V=1.5\times10^{-19}$, and $X_I=7.1\times10^{-27}$. The precise determination of these parameters enables us to measure the current noise and the corresponding Fano factor within 1 \% error.

Usually, the variation of the current fluctuation with the source-drain voltage $V_{\mbox{sd}}$ are analyzed with the  formula\cite{blanter}
\begin{equation}
S_I(V_{\mbox{sd}})=2FG\left[eV_{\mbox{sd}}\coth\left(\frac{eV_{\mbox{sd}}}{2k_BT_e}\right)-2k_BT_e\right]+4k_BT_eG,
\label{shotBB}
\end{equation}
where $T_e$ is electron temperature which is precisely determined by the thermal noise measurement. This formula takes the crossover between the thermal noise and the shot noise ($|eV_{\mbox{sd}}|\sim2k_BT_e$) into account. However, this method could overestimate the Fano factor when the current noise is affected by the $1/f$ noise in addition to the shot noise. In such a situation, the current noise is empirically known to be proportional to $V_{\mbox{sd}}^{\alpha}\quad (\alpha \sim 2)$ in the range $eV\gg2k_BT$. The data where a $1/f$ noise contribution was dominant are not taken into account in our analysis. xFor the remaining data we evaluated the Fano factor by fitting
\begin{equation}
S_I(V_{\mbox{sd}}) =2e|I|F
\label{shot2eIF}
\end{equation}
 in the range $|eV_{\mbox{sd}}|\gtrsim2k_BT_e$. Note that Eq.~(\ref{shot2eIF}) corresponds to  Eq.~(\ref{shotBB}) in the high bias voltage region. We use  Eq.~(\ref{shot2eIF}) as we found that it gives a more reliable Fano factor because the expression is much simpler than Eq.~(\ref{shotBB}).

Figure 2 shows a few typical examples of the current noise and the differential conductance as a function of  $V_{\mbox{sd}}$. The results of the fitting are superposed. Note that the Fano factor evaluated from simple linear function is consistent with that from function (2) within the precision of 1\%. Also we note that the thermal noise (current noise at $V_{\mbox{sd}}=0$ mV) is subtracted from the current noise shown in this article.

\section{Results and Discussion}

\subsection{Effect of magnetic field}

\begin{figure}[t]
\includegraphics[width=.99\linewidth]{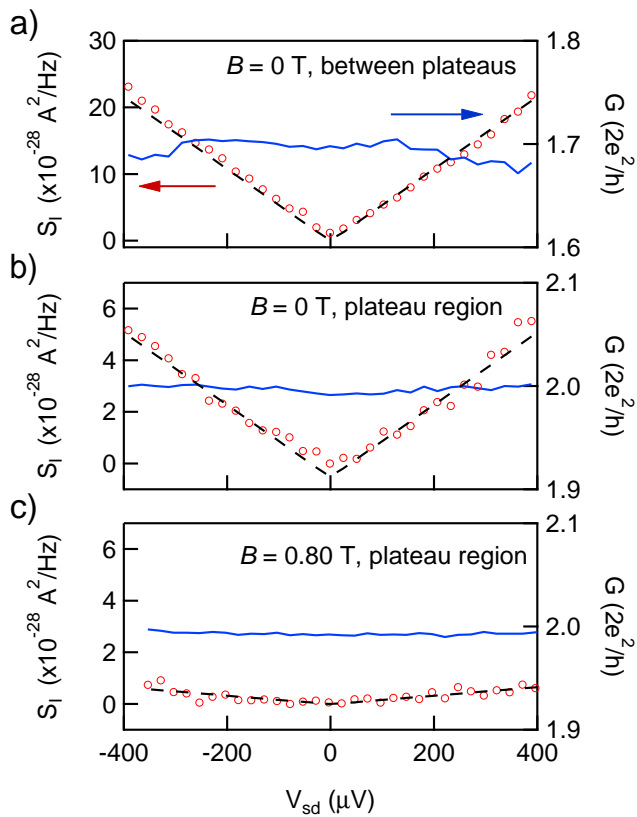}
\caption{\label{shotnoise_fig} $V_{\mbox{sd}}$ dependence of current noise (open circle) and conductance (solid line) of QPC 2 for different gate voltage and magnetic field. The dashed line shows fitting result. (a) In the region between plateaus, $G\sim1.7\times(2e^2/h)$ for zero magnetic field. (b) At the second plateau for zero magnetic field. (c) At the second plateau for $B=0.80$ T. The fitted Fano factor is $F=0.12$, 0.029, and $1.7\times10^{-3}$, respectively.}
\end{figure}

\begin{figure}[t]
\includegraphics[width=.99\linewidth]{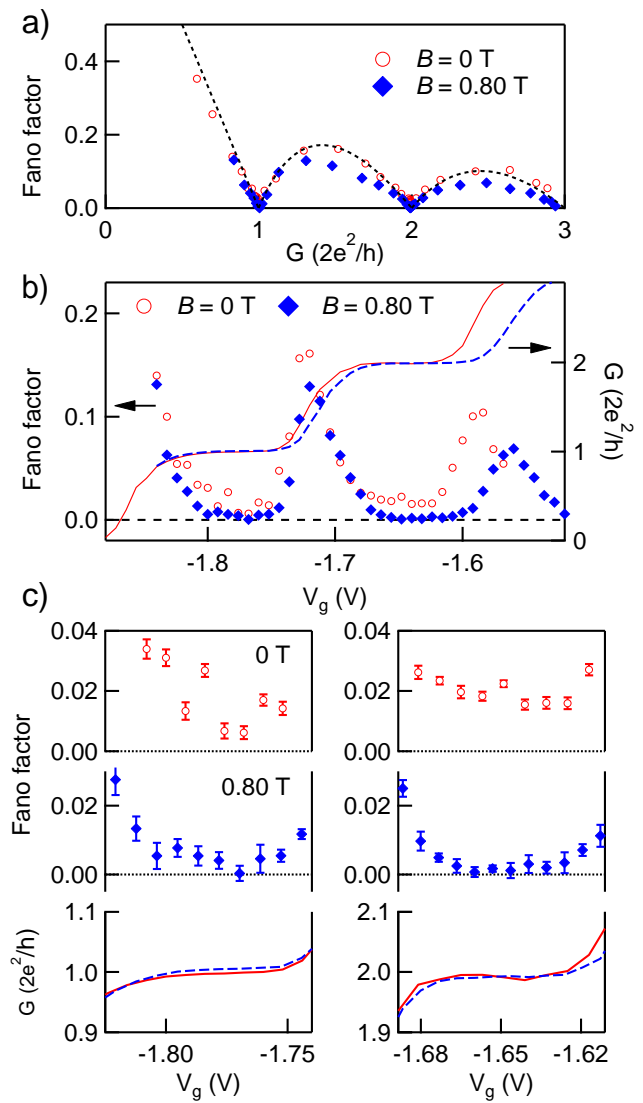}
\caption{\label{Fano_detail}a) Conductance dependence of the Fano factor for $B=0$ T (open circle) and 0.80 T (filled square). The black dashed line shows theoretical value without Zeeman splitting. The current noise at $G\sim1.7\times(2e^2/h)$ and $G=2\times(2e^2/h)$ is shown in Figs. 2(b) and 2(c), respectively. (b) $V_g$ dependence of Fano factor. The solid and dashed line curves conductance at $B$=0 T and 0.80 T, respectively. (c) Detailed plot of Fano factor and conductance against $V_g$ near the region of first plateau (left figure) and second plateau (right figure). The upper and middle panels show the Fano factors at  $B$=0 T and 0.80 T, respectively. The bottom panels show the conductance.  The solid and dashed curves represent conductance for $B$=0 T and 0.80 T, respectively. Note that the horizontal axis for 0.80 T data is rescaled to superpose the 0 T data.}
\end{figure}

We first investigate how QPC shot noise varies when a perpendicular magnetic field is applied to the 2DEG. 
Figures 3(a) and 3(b) show the Fano factor of  QPC 2 at 15 mK as a function of the conductance $G$ and the gate voltage $V_g$, respectively. The obtained Fano factors oscillate as the conductance increases from zero to $3\times(2e^2/h)$ and show a minimum at every conductance plateau. The dashed curve in Fig. 3 (a) represents the Fano factor  expected from the formula, $F=\sum_n T_ n(1-T_n)/\sum _n T_n$ and $G=2e^2/h\sum_n T_n$. The evaluated values seem to overall agree with the conventional theory\cite{blanter,buttiker1}. For example, as shown in Fig. 2 (a), the Fano factor is $F=0.121\pm0.003$ at $G=1.70\times(2e^2/h)$, which agrees well with the expected $F=0.124$ for $\sum _nT_n=1.7$.

However, it is important to notice that the Fano factor at the conductance plateau is slightly enhanced for zero magnetic field and suppressed by the magnetic field. Actually, at zero magnetic field, the current noise linearly increases with source-drain voltage even when conductance is an integer multiple of $2e^2/h$. Figure 2(b) shows a typical example. Here the quantized conductance of $2e^2/h$ is realized while the current noise shows a shot-noise-like V-shaped behavior. The Fano factor deduced from this noise data is $0.029\pm0.01$. It is noted that the differential conductance is perfectly flat and independent of $V_{\mbox{sd}}$. Therefore, the observed finite shot noise is not due to a non-linear effect. Figure 3(b) shows the evolution of the Fano factor of QPC as a function of $V_g$. The Fano factor obtained at $B=0$ T is finite even at the conductance plateaus (see the data plotted by open circles).

Interestingly, this phenomenon is greatly affected by applying a magnetic field perpendicular to the 2DEG. Figure 2(c) shows the current noise at the magnetic field $B=0.80$ T. The differential conductance is still independent of $V_{\mbox{sd}}$ as was the case in Fig. 2 b), while the noise is now largely suppressed almost to zero. From a comparison between Figs. 2(b) and 2(c), it is clear that the noise at conductance plateaus disappears for finite $B$ and results in $F=1.7\times10^{-3}$. The left and right panels of Fig. 3(c) show the expanded view to present the magnetic field effect on the Fano factor around the first and the second plateau region, respectively. The Fano factor is clearly finite at $B=0$ T against the expectation from the theory but reaches close to zero (typically less than $5\times10^{-3}$) at $B=0.80$ T as theory teaches us. The Fano factor at plateau region decreases almost monotonically as $B$ increases from 0 to 0.8 T [also see Figs. 6(b) and 6(c)].

At closer inspection of the conductance data at 0~T shown in the left bottom-most panel in Fig. 3(c), it deviates slightly from unity on the expanded scale of the left axis. The conductance gradually varies between 0.994 and 1.005 in unit of $2e^2/h$ in the middle of the plateau. This might suggest some finite scattering even in the middle of the plateau. However, the deviation is smaller than 1 \% and this alone cannot explain the obtained Fano factor which is as large as 0.01--0.03. For example, when the conductance is $0.994(2e^2/h)$ at $V_g =-1.784$~V, the expected Fano factor is $1-0.994 = 0.006$, which is much smaller than the observed value 0.027. In the same way, a slight non-monotonic behavior of the conductance at the plateau presented in the right-bottom panel in Fig. 3(c) does not explain the observed finite Fano factor of 0.02.

Here we make a brief comment on the spin polarization deduced by the Fano factor, although this is not the main point of the present work. The Fano factor between the neighboring plateaus decreases as $B$ increases from 0 T to 0.80 T; for example, at $G=1.5\times(2e^2/h)$, the Fano factor is $F=0.16$ at $B=0$ T, while $F=0.13$ at 0.80 T as shown in Fig. 3 b). This is attributed to the spin-resolved electron transport by the Zeeman splitting. If we simply assume that the up and down spin have a different constant transmission\cite{roche,dicarlo,reilly,nakamura}, the channel asymmetry at $B$=0.80 T is found to be 68 \% at most.

The reduction of the Fano factor was further investigated in the quantum Hall regime. Figure 4 a) shows the Fano factor and QPC conductance at $B=3.55$ T (filling factor=4) in the left and right axis as a function of $V_g$, respectively. When $V_g$ is swept from $-1.1$ V to $-1.5$ V, the conductance changes from the spin-resolved plateau $3\times(e^2/h)$ to the plateau $2e^2/h$. The Fano factor is very close to zero in these two plateaus and is finite only between them. The dashed curve shows the Fano factors deduced from the measured conductance, which agrees well with the obtained ones. Figure 4 b) shows the conductance and the current noise obtained at the middle of the plateau ($V_g=-1.35$ V). The current noise is almost zero and the conductance is totally independent of $V_{\mbox{sd}}$, in agreement with our naive expectation. Indeed, the Fano factor averaged over the plateau is $-2.2 \pm 2.5\times10^{-3}$, which statistically equals to zero. The perfect absence of the Fano factor directly reflects the dissipationless nature of the edge states. 

We also measured the in-plane magnetic field dependence of the Fano factor up to 0.8 T, and  found that it is independent of the magnetic field; the finite Fano factor obtained at zero field remains constant in the parallel magnetic fields up to 0.8 T. This observation, which is in clear contrast with that in perpendicular magnetic field, strongly suggests that the electron spin is not responsible for this phenomenon and that the perpendicular magnetic field  affects the Fano factor by influencing the electron motion confined in the 2DEG through the Lorentz force. The cyclotron radius for this 2DEG is, for example, 160 nm at 0.6 T and 120 nm at 0.8 T, which are of the same order of magnitude as the QPC width that conducting electrons feel. This may coincide with the fact that
the Fano factor is robustly suppressed when the field reaches these values [also see  Figs. 6(b) and 6(c)]. Thus, the semi-classical electron motion is very likely to be relevant in our observation. 

\begin{figure}[htb]
\includegraphics[width=.99\linewidth]{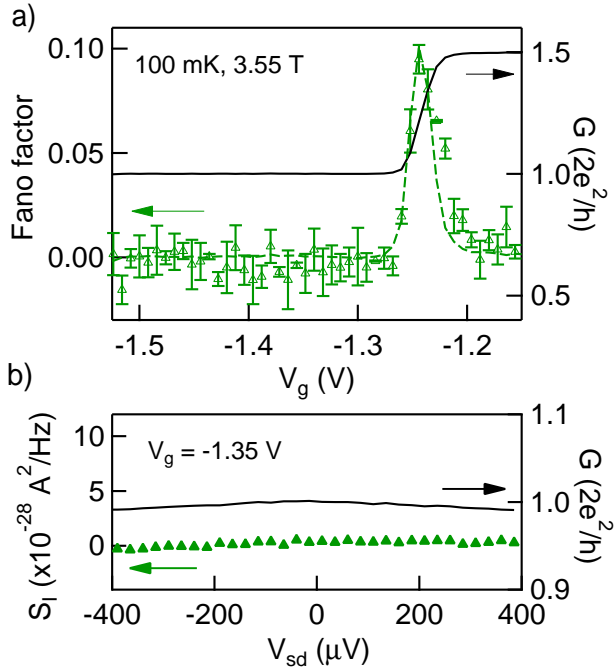}
\caption{\label{qhe} (a) $V_g$ dependence of Fano factor (triangle) and conductance (solid line) for the quantum Hall regime. The dashed curve shows the theoretical Fano factor deduced from the measured conductance. (b) Current noise (triangle) and conductance (solid line) as a function of $V_{\mbox{sd}}$ at the conductance plateau ($V_g=-1.35$ V).}
\end{figure}

\subsection{Fano factor at higher conductance}
We next focus on the shot noise at higher conductance plateaus. The behavior of the Fano factor as a function of the conductance up to $G=10\times(2e^2/h)$ is shown in Fig. 5(a). The expected Fano factor oscillation is obtained, which is overall consistent with the theory calculated from the conductance (shown by the dashed curve). Note that such a comparison between the theory and the experiment is only possible in the combination of a high quality QPC with many conductance plateaus and a precise noise measurement setup. Actually, the result shown in Fig. 5(a) nicely exemplifies the validity of the Landauer-B{\"u}ttiker formalism extending to the shot noise. Now we look carefully at the Fano factor at the plateaus. Here, the Fano factor averaged over the conductance plateau at each plateau region is summarized, which confirmed finite value in the range of 95 \% confidence interval [see Figs. 5(b) and 5(c)]. The Fano factor, which should be zero theoretically, becomes larger at higher conductance.

\begin{figure}[htb]
\includegraphics[clip,width=.99\linewidth]{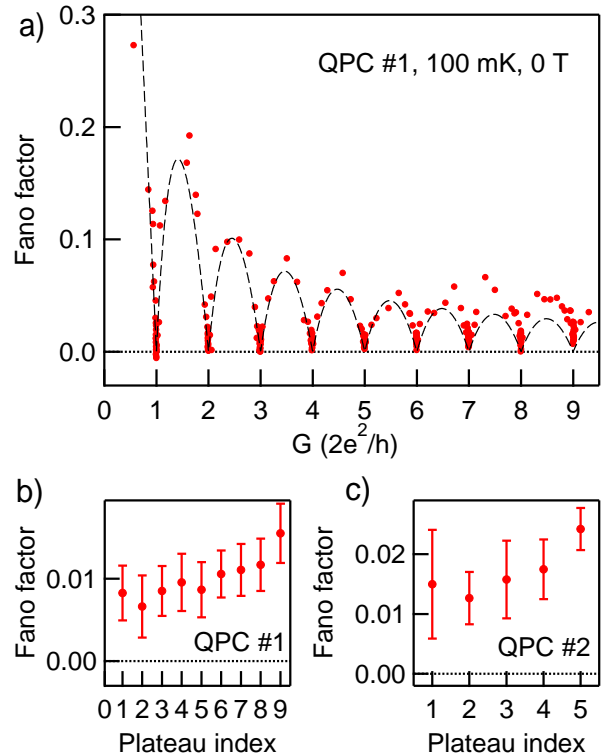}
\caption{\label{plateu}a) The Fano factor versus conductance of QPC 1 at zero magnetic field. The dashed curve shows the theoretical Fano factor deduced from the measured conductance. (b) and (c) The average value of Fano factor at each conductance plateau for QPC 1 (left figure) and QPC 2 (right figure).}
\end{figure}

\subsection{Electron heating model}

So far, we have experimentally established that the Fano factor is finite at the conductance plateau and is reduced by applied perpendicular magnetic field.  Previous research already observed this phenomenon and explained it by electron heating~\cite{kumar}. As we show below, also in our case, the electron heating model seems to work quantitatively. Now we show a systematic analysis of our observations based on this model and discuss its implication.

As the energy dissipation does not occur exactly at QPC in the ballistic transport, it is not trivial how the electron heating affects the shot noise. For mesoscopic systems, the energy dissipation of electrons occurs due to electron-electron scattering and acoustic phonon emission by the injected electrons\cite{steinbach,dzurak}. Since the latter is negligible at very low temperature, electron thermalization mainly takes place via thermal conduction in the reservoirs. More specifically, the hot electrons injected into the QPC give rise to thermal dissipation only at the connection with 2DEG lead where a large number of conduction channels exist. Hence, there may exist nontrivial thermal noise generated by an increased temperature of electrons in the vicinity of the QPC. 

\begin{figure}[t]
\includegraphics[width=.99\linewidth]{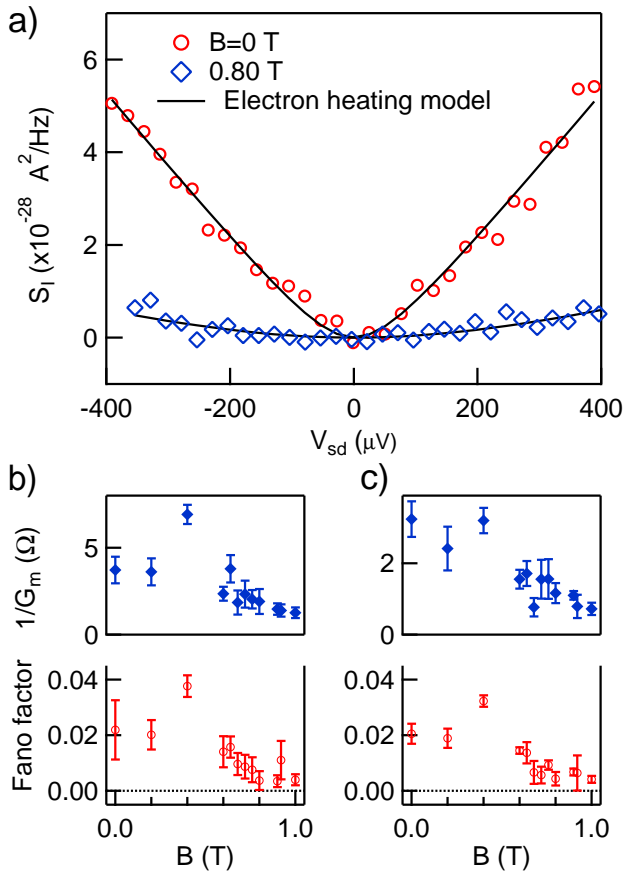}
\caption{\label{gmb}a) Current noise at 2nd plateau versus $V_{\mbox{sd}}$ for $B=0$ T (open circle) and $0.80$ T (open square). The solid curves show the fit of the data assuming $S_I(V_{\mbox{sd}})= 4k_B T_e^{JH}G$. $1/G_m$ is $2.78\ \Omega$ and $0.14\ \Omega$ for $B=0$ T and $0.80$ T, respectively. (b) The variation of the nominal Fano factor (red circle) and $1/G_m$ (blue square) with perpendicular magnetic field for the first and the second plateau region. These values are derived by fitting the measured current noise for QPC 2.}
\end{figure}

Because both charge and heat are transported by conducting electrons, we can relate the heat conductivity $\kappa$ to $G$ by Wiedemann-Franz law. Assuming one dimensional heat diffusion, Kumar \textit{et al}.\cite{kumar} showed that the formula to express the effective electron temperature $T_e^{JH}$, which the electrons feel in the lead is expressed by the relation,
\begin{equation}
\left(\frac{T_e^{JH}}{T_e}\right)^2=1+\frac{24}{\pi^2}\frac{G}{G_m}\left(1+\frac{2G}{G_m}\right)\left(\frac{eV_{\mbox{sd}}}{2k_BT_e}\right)^2
\label{HeatingT}
\end{equation}
where $G_m$ is the conductance of the 2DEG leads. If we can neglect the electron heating effect, we can take $1/G_m=0$ and $T_e^{JH}$ equals to $T_e$. On the other hand, for finite $1/G_m$, the hot electrons that have passed through the QPC heat up the lead. Thus, phenomenologically, $1/G_m$ characterizes the heat conduction associated with the electron-electron and/or electron-phonon interaction. By assuming that on the plateau $F$ exactly equals to zero and thus $S_I(V_{\mbox{sd}})= 4k_B T_e^{JH}G$, we can precisely determine the parameter $G_m$ without any difficulty. Because the conventional shot noise theory predicts that the Fano factor is zero and thus the additional noise can be simply attributed to the heating effect.

\begin{figure}[t]
\includegraphics[width=.9\linewidth]{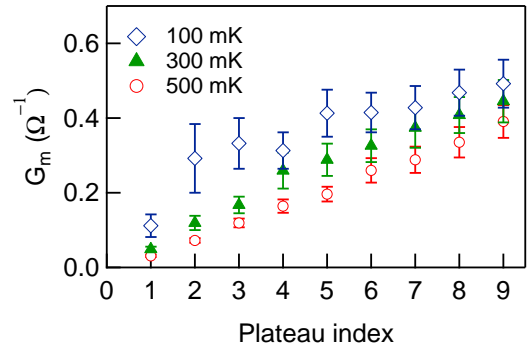}
\caption{\label{gmt}The conductance dependence of parameter $G_m$ of QPC 1 for $T_e$=500 mK (circle), 300 mK (triangle), and 100 mK (square). $G_m$'s are obtained at 1st, 2nd, ..., 9th plateau, respectively.}
\end{figure}

Figure 6(a) shows typical examples of the analysis of the current noise data at $B=0$ T and 0.80 T, which is the same ones as shown in Figs. 2(a) and 2(b). $1/G_m$ was obtained to be $2.78\pm0.14\ \Omega$ and $0.14\pm0.02\ \Omega$ for $B=0$ T and 0.80 T, respectively.

$B$ dependencies of the Fano factor and $1/G_m$ for the first and second plateaus are summarized in Figs. 6(b) and 6(c), respectively. Here we also show the Fano factor obtained through the conventional treatment. As expected, the Fano factor, which is nominally estimated, has a strong correlation with $1/G_m$.

As is evident from Figs. 6(b) and 6(c), $1/G_m$ in both plateau regions decreases as $B$ rises above 0.6 T and then it continues to decrease up to 0.9 T. Finally, $1/G_m$ is very close to zero in the quantum Hall regime as the shot noise is absent [see Fig. 4(a)]. This implies that electrons near the QPC are insensitive to the electron heating effect at high magnetic field. Indeed, source-drain voltage $V_{\mbox{sd}}=400\ \mu$V typically gives $T_{e}^{JH}-T_e=73\ \rm{mK}$ and 38 mK at the second plateau for $B=0$ T and 1.0 T, respectively.

It is meaningful to compare the present $1/G_m$ with those reported previously. Previous shot noise experiments for QPC fabricated on GaAs/AlGaAs ($\mu=200\ \rm{m^2/Vs}$)\cite{kumar} and InGaAs/InGaAsP ($\mu=11.65\ \rm{m^2/Vs}$)\cite{nishihara} heterostructure report $1/G_m\sim250\ \Omega$ and 110\ $\Omega$, respectively. In the present case, our result $1/G_m=2.78$ and 0.14 shows that the heating effect is very small compared to the previous experiments. Since our QPCs are fabricated on high-mobility 2DEG, the heat conduction is much more efficient than in the previous cases.

While we adopted a simple one-dimensional model to analyze our results, the observed suppression of the electron heating by the magnetic field prior to the edge channel formation is most probably related to the anisotropy of thermal conductivity in 2DEG. The experimental result on the in-plane magnetic field supports this idea as we discussed already. Naively we may speculate as follows: at zero magnetic field, the electrons injected from one reservoir to the QPC run straight to the other reservoir and electron kinetic energy is dissipated there. This may create a kind of ``hot spot'' in the vicinity of the QPC and therefore the finite noise is induced even at the conductance plateau. On the other hand, at finite magnetic field, the electron motion and thus the heat conduction has chirality because of the Lorentz force. The dissipated energy or the ``hot spot'' is carried away from the QPC efficiently by this chirality\cite{HirayamaJLTP2013}. This topic is nothing but the thermal Hall effect. However, we have to admit that the present one-dimensional model is not at all appropriate to address this situation quantitatively. Further theoretical attempt is preferable to treat the electron and heat conduction in the presence of a magnetic field.

Lastly we investigate the variation of parameter $G_m$ with the temperature for different conductances. Figure 7 gives the value of $G_m$ as a function of conductance for different temperature. One can notice that the parameter $G_m$ increases monotonically with $G$, which helps electron heat to diffuse. Thus, it may suggest that thermal relaxation is generated where electron system near QPC changes from one dimensional to two dimensional. When the QPC conductance increases, the coupling between the QPC and the 2DEG leads becomes strong and the heating effect becomes less significant.

\section{Conclusion}

To conclude, we confirm that excess noise observed in the QPC is closely associated with thermal relaxation nearby in the 2DEG.
This means that  it is not possible to ignore the electron heating effect  even in a high mobility device since it produces finite current noise in a perfectly transmitting system.

Moreover, electron heating effect absolutely vanishes in the quantum Hall regime. Since spatially-separated channels largely reduce electron interaction, relaxation of thermally excited conduction electrons take place far from QPC. Therefore, our results indicate that enhanced current noise arises from relaxation of electrons near the QPC. However, electron heating has been already suppressed before quantum Hall state is formed. We attribute the reason for the observed decrease in electron heating to temperature gradient in transverse direction of 2DEG created by perpendicular magnetic field.

Our experiments closely relate to the thermal Hall effect. Further experimental and theoretical work on the relation between the electron heating and magnetic field would serve to study heating flow in low-dimensional systems or extended electron waveguide circuits.

\section*{Acknowledgements}
We appreciate the fruitful discussion with Preden Roulleau and Yasuhiro Utsumi. This work was partially supported by a Grant-in-Aid for Scientific Research (S) (JP26220711), JSPS KAKENHI (JP15K17680), Invitation Fellowships for Research in Japan from JSPS, Grant-in-Aid for Scientific Research on Innovative Areas ``Fluctuation \& Structure'' (JP25103003) and ``Topological Materials Science'' (KAKENHI Grant JP15H05854), and Yazaki Memorials Foundation for Science and Technology. K. K. acknowledges the stimulating discussions in the meeting of the Cooperative Research Project of RIEC, Tohoku University.


\begin{thebibliography}{37}%
\makeatletter
\providecommand \@ifxundefined [1]{%
 \@ifx{#1\undefined}
}%
\providecommand \@ifnum [1]{%
 \ifnum #1\expandafter \@firstoftwo
 \else \expandafter \@secondoftwo
 \fi
}%
\providecommand \@ifx [1]{%
 \ifx #1\expandafter \@firstoftwo
 \else \expandafter \@secondoftwo
 \fi
}%
\providecommand \natexlab [1]{#1}%
\providecommand \enquote  [1]{``#1''}%
\providecommand \bibnamefont  [1]{#1}%
\providecommand \bibfnamefont [1]{#1}%
\providecommand \citenamefont [1]{#1}%
\providecommand \href@noop [0]{\@secondoftwo}%
\providecommand \href [0]{\begingroup \@sanitize@url \@href}%
\providecommand \@href[1]{\@@startlink{#1}\@@href}%
\providecommand \@@href[1]{\endgroup#1\@@endlink}%
\providecommand \@sanitize@url [0]{\catcode `\\12\catcode `\$12\catcode
  `\&12\catcode `\#12\catcode `\^12\catcode `\_12\catcode `\%12\relax}%
\providecommand \@@startlink[1]{}%
\providecommand \@@endlink[0]{}%
\providecommand \url  [0]{\begingroup\@sanitize@url \@url }%
\providecommand \@url [1]{\endgroup\@href {#1}{\urlprefix }}%
\providecommand \urlprefix  [0]{URL }%
\providecommand \Eprint [0]{\href }%
\providecommand \doibase [0]{http://dx.doi.org/}%
\providecommand \selectlanguage [0]{\@gobble}%
\providecommand \bibinfo  [0]{\@secondoftwo}%
\providecommand \bibfield  [0]{\@secondoftwo}%
\providecommand \translation [1]{[#1]}%
\providecommand \BibitemOpen [0]{}%
\providecommand \bibitemStop [0]{}%
\providecommand \bibitemNoStop [0]{.\EOS\space}%
\providecommand \EOS [0]{\spacefactor3000\relax}%
\providecommand \BibitemShut  [1]{\csname bibitem#1\endcsname}%
\let\auto@bib@innerbib\@empty
\bibitem [{\citenamefont {van Wees}\ \emph {et~al.}(1988)\citenamefont {van
  Wees}, \citenamefont {van Houten}, \citenamefont {Beenakker}, \citenamefont
  {Williamson}, \citenamefont {Kouwenhoven}, \citenamefont {van~der Marel},\
  and\ \citenamefont {Foxon}}]{vanwees}%
  \BibitemOpen
  \bibfield  {author} {\bibinfo {author} {\bibfnamefont {B.~J.}\ \bibnamefont
  {van Wees}}, \bibinfo {author} {\bibfnamefont {H.}~\bibnamefont {van
  Houten}}, \bibinfo {author} {\bibfnamefont {C.~W.~J.}\ \bibnamefont
  {Beenakker}}, \bibinfo {author} {\bibfnamefont {J.~G.}\ \bibnamefont
  {Williamson}}, \bibinfo {author} {\bibfnamefont {L.~P.}\ \bibnamefont
  {Kouwenhoven}}, \bibinfo {author} {\bibfnamefont {D.}~\bibnamefont {van~der
  Marel}}, and \bibinfo {author} {\bibfnamefont {C.~T.}\ \bibnamefont
  {Foxon}},\ }\href@noop {} {\bibfield  {journal} {\bibinfo  {journal} {Phys.\
  Rev.\ Lett.}\ }\textbf {\bibinfo {volume} {60}},\ \bibinfo {pages} {848}
  (\bibinfo {year} {1988})}\BibitemShut {NoStop}%
\bibitem [{\citenamefont {{D. A. Wharam \textit{et al}.}}(1988)}]{wharam}%
  \BibitemOpen
  \bibfield  {author} {\bibinfo {author} {\bibnamefont {{D. A. Wharam
  \textit{et al}.}}},\ }\href@noop {} {\bibfield  {journal} {\bibinfo
  {journal} {J. Phys. C}\ }\textbf {\bibinfo {volume} {21}},\ \bibinfo {pages}
  {L209} (\bibinfo {year} {1988})}\BibitemShut {NoStop}%
\bibitem [{\citenamefont {Thomas}\ \emph {et~al.}(1996)\citenamefont {Thomas},
  \citenamefont {Nicholls}, \citenamefont {Simmons}, \citenamefont {Pepper},
  \citenamefont {Mace}, and \citenamefont {Ritchie}}]{thomas1}%
  \BibitemOpen
  \bibfield  {author} {\bibinfo {author} {\bibfnamefont {K.~J.}\ \bibnamefont
  {Thomas}}, \bibinfo {author} {\bibfnamefont {J.~T.}\ \bibnamefont
  {Nicholls}}, \bibinfo {author} {\bibfnamefont {M.~Y.}\ \bibnamefont
  {Simmons}}, \bibinfo {author} {\bibfnamefont {M.}~\bibnamefont {Pepper}},
  \bibinfo {author} {\bibfnamefont {D.~R.}\ \bibnamefont {Mace}}, and
  \bibinfo {author} {\bibfnamefont {D.~A.}\ \bibnamefont {Ritchie}},\
  }\href@noop {} {\bibfield  {journal} {\bibinfo  {journal} {Phys.\ Rev.\
  Lett.}\ }\textbf {\bibinfo {volume} {77}},\ \bibinfo {pages} {135} (\bibinfo
  {year} {1996})}\BibitemShut {NoStop}%
\bibitem [{\citenamefont {Thomas}\ \emph
  {et~al.}(2000{\natexlab{a}})\citenamefont {Thomas}, \citenamefont {Nicholls},
  \citenamefont {Appleyard}, \citenamefont {Simmons}, \citenamefont {Pepper},
  \citenamefont {Mace}, \citenamefont {Tribe}, and \citenamefont
  {Ritchie}}]{thomas2}%
  \BibitemOpen
  \bibfield  {author} {\bibinfo {author} {\bibfnamefont {K.~J.}\ \bibnamefont
  {Thomas}}, \bibinfo {author} {\bibfnamefont {J.~T.}\ \bibnamefont
  {Nicholls}}, \bibinfo {author} {\bibfnamefont {N.~J.}\ \bibnamefont
  {Appleyard}}, \bibinfo {author} {\bibfnamefont {M.~Y.}\ \bibnamefont
  {Simmons}}, \bibinfo {author} {\bibfnamefont {M.}~\bibnamefont {Pepper}},
  \bibinfo {author} {\bibfnamefont {D.~R.}\ \bibnamefont {Mace}}, \bibinfo
  {author} {\bibfnamefont {W.~R.}\ \bibnamefont {Tribe}},  and \bibinfo
  {author} {\bibfnamefont {D.~A.}\ \bibnamefont {Ritchie}},\ }\href@noop {}
  {\bibfield  {journal} {\bibinfo  {journal} {Phys. Rev. B}\ }\textbf {\bibinfo
  {volume} {58}},\ \bibinfo {pages} {4846} (\bibinfo {year}
  {1998}{\natexlab{a}})}\BibitemShut {NoStop}%
\bibitem [{\citenamefont {Thomas}\ \emph
  {et~al.}(2000{\natexlab{b}})\citenamefont {Thomas}, \citenamefont {Nicholls},
  \citenamefont {Pepper}, \citenamefont {Tribe}, \citenamefont {Simmons},\ and\
  \citenamefont {Ritchie}}]{thomas3}%
  \BibitemOpen
  \bibfield  {author} {\bibinfo {author} {\bibfnamefont {K.~J.}\ \bibnamefont
  {Thomas}}, \bibinfo {author} {\bibfnamefont {J.~T.}\ \bibnamefont
  {Nicholls}}, \bibinfo {author} {\bibfnamefont {M.}~\bibnamefont {Pepper}},
  \bibinfo {author} {\bibfnamefont {W.~R.}\ \bibnamefont {Tribe}}, \bibinfo
  {author} {\bibfnamefont {M.~Y.}\ \bibnamefont {Simmons}}, \ and\ \bibinfo
  {author} {\bibfnamefont {D.~A.}\ \bibnamefont {Ritchie}},\ }\href@noop {}
  {\bibfield  {journal} {\bibinfo  {journal} {Phys. Rev. B}\ }\textbf {\bibinfo
  {volume} {61}},\ \bibinfo {pages} {R13365} (\bibinfo {year}
  {2000}{\natexlab{b}})}\BibitemShut {NoStop}%
\bibitem [{\citenamefont {Nuttinck}\ \emph {et~al.}(2000)\citenamefont
  {Nuttinck}, \citenamefont {Hashimoto}, \citenamefont {Miyashita},
  \citenamefont {Saku}, \citenamefont {Yamamoto},\ and\ \citenamefont
  {Hirayama}}]{nuttinck}%
  \BibitemOpen
  \bibfield  {author} {\bibinfo {author} {\bibfnamefont {S.}~\bibnamefont
  {Nuttinck}}, \bibinfo {author} {\bibfnamefont {K.}~\bibnamefont {Hashimoto}},
  \bibinfo {author} {\bibfnamefont {S.}~\bibnamefont {Miyashita}}, \bibinfo
  {author} {\bibfnamefont {T.}~\bibnamefont {Saku}}, \bibinfo {author}
  {\bibfnamefont {Y.}~\bibnamefont {Yamamoto}}, and \bibinfo {author}
  {\bibfnamefont {Y.}~\bibnamefont {Hirayama}},\ }\href@noop {} {\bibfield
  {journal} {\bibinfo  {journal} {Jpn. J. Appl. Phys}\ }\textbf {\bibinfo
  {volume} {39}},\ \bibinfo {pages} {L655} (\bibinfo {year}
  {2000})}\BibitemShut {NoStop}%
\bibitem [{\citenamefont {Reilly}\ \emph {et~al.}(2002)\citenamefont {Reilly},
  \citenamefont {Buehler}, \citenamefont {OBrien}, \citenamefont {Hamilton},
  \citenamefont {Dzurak}, \citenamefont {Clark}, \citenamefont {Kane},
  \citenamefont {Pfeiffer}, and \citenamefont {West}}]{reilly}%
  \BibitemOpen
  \bibfield  {author} {\bibinfo {author} {\bibfnamefont {D.~J.}\ \bibnamefont
  {Reilly}}, \bibinfo {author} {\bibfnamefont {T.~M.}\ \bibnamefont {Buehler}},
  \bibinfo {author} {\bibfnamefont {J.~L.}\ \bibnamefont {OBrien}}, \bibinfo
  {author} {\bibfnamefont {A.~R.}\ \bibnamefont {Hamilton}}, \bibinfo {author}
  {\bibfnamefont {A.~S.}\ \bibnamefont {Dzurak}}, \bibinfo {author}
  {\bibfnamefont {R.~G.}\ \bibnamefont {Clark}}, \bibinfo {author}
  {\bibfnamefont {B.~E.}\ \bibnamefont {Kane}}, \bibinfo {author}
  {\bibfnamefont {L.~N.}\ \bibnamefont {Pfeiffer}}, and \bibinfo {author}
  {\bibfnamefont {K.~W.}\ \bibnamefont {West}},\ }\href@noop {} {\bibfield
  {journal} {\bibinfo  {journal} {Phys.\ Rev.\ Lett.}\ }\textbf {\bibinfo
  {volume} {89}},\ \bibinfo {pages} {246801} (\bibinfo {year}
  {2002})}\BibitemShut {NoStop}%
\bibitem [{\citenamefont {Crook}\ \emph {et~al.}(2006)\citenamefont {Crook},
  \citenamefont {Prance}, \citenamefont {Thomas}, \citenamefont {Chorley},
  \citenamefont {Farrer}, \citenamefont {Ritchie}, \citenamefont {Pepper},\
  and\ \citenamefont {Smith}}]{crook}%
  \BibitemOpen
  \bibfield  {author} {\bibinfo {author} {\bibfnamefont {R.}~\bibnamefont
  {Crook}}, \bibinfo {author} {\bibfnamefont {J.}~\bibnamefont {Prance}},
  \bibinfo {author} {\bibfnamefont {K.~J.}\ \bibnamefont {Thomas}}, \bibinfo
  {author} {\bibfnamefont {S.~J.}\ \bibnamefont {Chorley}}, \bibinfo {author}
  {\bibfnamefont {I.}~\bibnamefont {Farrer}}, \bibinfo {author} {\bibfnamefont
  {D.~A.}\ \bibnamefont {Ritchie}}, \bibinfo {author} {\bibfnamefont
  {M.}~\bibnamefont {Pepper}}, and \bibinfo {author} {\bibfnamefont {C.~G.}\
  \bibnamefont {Smith}},\ }\href@noop {} {\bibfield  {journal} {\bibinfo
  {journal} {Science}\ }\textbf {\bibinfo {volume} {312}},\ \bibinfo {pages}
  {1359} (\bibinfo {year} {2006})}\BibitemShut {NoStop}%
\bibitem [{\citenamefont {Yoon}\ \emph {et~al.}(2007)\citenamefont {Yoon},
  \citenamefont {Mourokh}, \citenamefont {Morimoto}, \citenamefont {Aoki},
  \citenamefont {Ochiai}, \citenamefont {Reno},\ and\ \citenamefont
  {Bird}}]{yoon}%
  \BibitemOpen
  \bibfield  {author} {\bibinfo {author} {\bibfnamefont {Y.}~\bibnamefont
  {Yoon}}, \bibinfo {author} {\bibfnamefont {L.}~\bibnamefont {Mourokh}},
  \bibinfo {author} {\bibfnamefont {T.}~\bibnamefont {Morimoto}}, \bibinfo
  {author} {\bibfnamefont {N.}~\bibnamefont {Aok}}, \bibinfo {author}
  {\bibfnamefont {Y.}~\bibnamefont {Ochiai}}, \bibinfo {author} {\bibfnamefont
  {J.~L.}\ \bibnamefont {Reno}}, and \bibinfo {author} {\bibfnamefont
  {J.~P.}\ \bibnamefont {Bird}},\ }\href@noop {} {\bibfield  {journal}
  {\bibinfo  {journal} {Phys.\ Rev.\ Lett.}\ }\textbf {\bibinfo {volume}
  {99}},\ \bibinfo {pages} {136805} (\bibinfo {year} {2007})}\BibitemShut
  {NoStop}%
\bibitem [{\citenamefont {Chung}\ \emph {et~al.}(2007)\citenamefont {Chung},
  \citenamefont {Jo}, \citenamefont {Chang}, \citenamefont {Lee}, \citenamefont
  {Zaffalon}, \citenamefont {Umansky},\ and\ \citenamefont {Heiblum}}]{chung}%
  \BibitemOpen
  \bibfield  {author} {\bibinfo {author} {\bibfnamefont {Y.}~\bibnamefont
  {Chung}}, \bibinfo {author} {\bibfnamefont {S.}~\bibnamefont {Jo}}, \bibinfo
  {author} {\bibfnamefont {D.~I.}\ \bibnamefont {Chang}}, \bibinfo {author}
  {\bibfnamefont {H.~J.}\ \bibnamefont {Lee}}, \bibinfo {author} {\bibfnamefont
  {M.}~\bibnamefont {Zaffalon}}, \bibinfo {author} {\bibfnamefont
  {V.}~\bibnamefont {Umansky}}, and \bibinfo {author} {\bibfnamefont
  {M.}~\bibnamefont {Heiblum}},\ }\href@noop {} {\bibfield  {journal} {\bibinfo
   {journal} {Phys.\ Rev.\ B}\ }\textbf {\bibinfo {volume} {76}},\ \bibinfo
  {pages} {035316} (\bibinfo {year} {2007})}\BibitemShut {NoStop}%
\bibitem [{\citenamefont {{A. Kristensen \textit{et al}.}}(2000)}]{kritensen}%
  \BibitemOpen
  \bibfield  {author} {\bibinfo {author} {\bibnamefont {{A. Kristensen
  \textit{et al}.}}},\ }\href@noop {} {\bibfield  {journal} {\bibinfo
  {journal} {Phys.\ Rev.\ B}\ }\textbf {\bibinfo {volume} {62}},\ \bibinfo
  {pages} {10950} (\bibinfo {year} {2000})}\BibitemShut {NoStop}%
\bibitem [{\citenamefont {Cronenwett}\ \emph {et~al.}(2002)\citenamefont
  {Cronenwett}, \citenamefont {Lynch}, \citenamefont {Goldhaber-Gordon},
  \citenamefont {Kouwenhoven}, \citenamefont {Marcus}, \citenamefont {Hirose},
  \citenamefont {Wingreen},\ and\ \citenamefont {Umansky}}]{cronenwett}%
  \BibitemOpen
  \bibfield  {author} {\bibinfo {author} {\bibfnamefont {S.~M.}\ \bibnamefont
  {Cronenwett}}, \bibinfo {author} {\bibfnamefont {H.~J.}\ \bibnamefont
  {Lynch}}, \bibinfo {author} {\bibfnamefont {D.}~\bibnamefont
  {Goldhaber-Gordon}}, \bibinfo {author} {\bibfnamefont {L.~P.}\ \bibnamefont
  {Kouwenhoven}}, \bibinfo {author} {\bibfnamefont {C.~M.}\ \bibnamefont
  {Marcus}}, \bibinfo {author} {\bibfnamefont {K.}~\bibnamefont {Hirose}},
  \bibinfo {author} {\bibfnamefont {N.~S.}\ \bibnamefont {Wingreen}}, and
  \bibinfo {author} {\bibfnamefont {V.}~\bibnamefont {Umansky}},\ }\href@noop
  {} {\bibfield  {journal} {\bibinfo  {journal} {Phys.\ Rev.\ Lett.}\ }\textbf
  {\bibinfo {volume} {88}},\ \bibinfo {pages} {226805} (\bibinfo {year}
  {2002})}\BibitemShut {NoStop}%
\bibitem [{\citenamefont {Lesovik}(1989)}]{lesovik}%
  \BibitemOpen
  \bibfield  {author} {\bibinfo {author} {\bibfnamefont {G.~B.}\ \bibnamefont
  {Lesovik}},\ }\href@noop {} {\bibfield  {journal} {\bibinfo  {journal}
  {JETP}\ }\textbf {\bibinfo {volume} {49}},\ \bibinfo {pages} {592} (\bibinfo
  {year} {1989})}\BibitemShut {NoStop}%
\bibitem [{\citenamefont {B{\"u}ttiker}(1990{\natexlab{a}})}]{buttiker1}%
  \BibitemOpen
  \bibfield  {author} {\bibinfo {author} {\bibfnamefont {M.}~\bibnamefont
  {B{\"u}ttiker}},\ }\href@noop {} {\bibfield  {journal} {\bibinfo  {journal}
  {Phys.\ Rev.\ Lett.}\ }\textbf {\bibinfo {volume} {65}},\ \bibinfo {pages}
  {2901} (\bibinfo {year} {1990}{\natexlab{a}})}\BibitemShut {NoStop}%
\bibitem [{\citenamefont {B{\"u}ttiker}(1992)}]{buttiker2}%
  \BibitemOpen
  \bibfield  {author} {\bibinfo {author} {\bibfnamefont {M.}~\bibnamefont
  {B{\"u}ttiker}},\ }\href@noop {} {\bibfield  {journal} {\bibinfo  {journal}
  {Phys.\ Rev.\ B}\ }\textbf {\bibinfo {volume} {46}},\ \bibinfo {pages}
  {12485} (\bibinfo {year} {1992})}\BibitemShut {NoStop}%
\bibitem [{\citenamefont {{Th. Martin and R. Landauer}}(1992)}]{landauer}%
  \BibitemOpen
  \bibfield  {author} {\bibinfo {author} {\bibnamefont {{Th. Martin and R.
  Landauer}}},\ }\href@noop {} {\bibfield  {journal} {\bibinfo  {journal}
  {Phys.\ Rev.\ B}\ }\textbf {\bibinfo {volume} {45}},\ \bibinfo {pages} {1742}
  (\bibinfo {year} {1992})}\BibitemShut {NoStop}%
\bibitem [{\citenamefont {B{\"u}ttiker}(1990{\natexlab{b}})}]{buttiker3}%
  \BibitemOpen
  \bibfield  {author} {\bibinfo {author} {\bibfnamefont {M.}~\bibnamefont
  {B{\"u}ttiker}},\ }\href@noop {} {\bibfield  {journal} {\bibinfo  {journal}
  {Phys.\ Rev.\ B}\ }\textbf {\bibinfo {volume} {41}},\ \bibinfo {pages} {7906}
  (\bibinfo {year} {1990}{\natexlab{b}})}\BibitemShut {NoStop}%
\bibitem [{\citenamefont {Blanter}\ and\ \citenamefont
  {Buttiker}(2000)}]{blanter}%
  \BibitemOpen
  \bibfield  {author} {\bibinfo {author} {\bibfnamefont {Y.~M.}\ \bibnamefont
  {Blanter}}\ and\ \bibinfo {author} {\bibfnamefont {M.}~\bibnamefont
  {B{\"u}ttiker}},\ }\href@noop {} {\bibfield  {journal} {\bibinfo  {journal}
  {Phys.\ Rep}\ }\textbf {\bibinfo {volume} {336}},\ \bibinfo {pages} {1}
  (\bibinfo {year} {2000})}\BibitemShut {NoStop}%
\bibitem [{\citenamefont {Zarchin}\ \emph {et~al.}(2008)\citenamefont
  {Zarchin}, \citenamefont {Zaffalon}, \citenamefont {Heiblum}, \citenamefont
  {Mahalu},\ and\ \citenamefont {Umansky}}]{PhysRevB.77.241303}%
  \BibitemOpen
  \bibfield  {author} {\bibinfo {author} {\bibfnamefont {O.}~\bibnamefont
  {Zarchin}}, \bibinfo {author} {\bibfnamefont {M.}~\bibnamefont {Zaffalon}},
  \bibinfo {author} {\bibfnamefont {M.}~\bibnamefont {Heiblum}}, \bibinfo
  {author} {\bibfnamefont {D.}~\bibnamefont {Mahalu}}, and \bibinfo {author}
  {\bibfnamefont {V.}~\bibnamefont {Umansky}},\ }\href {\doibase
  10.1103/PhysRevB.77.241303} {\bibfield  {journal} {\bibinfo  {journal} {Phys.
  Rev. B}\ }\textbf {\bibinfo {volume} {77}},\ \bibinfo {pages} {241303}
  (\bibinfo {year} {2008})}\BibitemShut {NoStop}%
\bibitem [{\citenamefont {Delattre}\ \emph {et~al.}(2009)\citenamefont
  {Delattre}, \citenamefont {Feuillet-Palma}, \citenamefont {Herrmann},
  \citenamefont {Morfin}, \citenamefont {Berroir}, \citenamefont {Feve},
  \citenamefont {Placais}, \citenamefont {Glattli}, \citenamefont {Choi},
  \citenamefont {Mora},\ and\ \citenamefont {Kontos}}]{DelattreNatPhys2009}%
  \BibitemOpen
  \bibfield  {author} {\bibinfo {author} {\bibfnamefont {T.}~\bibnamefont
  {Delattre}}, \bibinfo {author} {\bibfnamefont {C.}~\bibnamefont
  {Feuillet-Palma}}, \bibinfo {author} {\bibfnamefont {L.~G.}\ \bibnamefont
  {Herrmann}}, \bibinfo {author} {\bibfnamefont {P.}~\bibnamefont {Morfin}},
  \bibinfo {author} {\bibfnamefont {J.-M.}\ \bibnamefont {Berroir}}, \bibinfo
  {author} {\bibfnamefont {G.}~\bibnamefont {Feve}}, \bibinfo {author}
  {\bibfnamefont {B.}~\bibnamefont {Placais}}, \bibinfo {author} {\bibfnamefont
  {D.~C.}\ \bibnamefont {Glattli}}, \bibinfo {author} {\bibfnamefont {M.-S.}\
  \bibnamefont {Choi}}, \bibinfo {author} {\bibfnamefont {C.}~\bibnamefont
  {Mora}}, and \bibinfo {author} {\bibfnamefont {T.}~\bibnamefont
  {Kontos}},\ }\href {\doibase 10.1038/nphys1186} {\bibfield  {journal}
  {\bibinfo  {journal} {Nat. Phys.}\ }\textbf {\bibinfo {volume} {5}},\
  \bibinfo {pages} {208} (\bibinfo {year} {2009})}\BibitemShut {NoStop}%
\bibitem [{\citenamefont {Yamauchi}\ \emph {et~al.}(2011)\citenamefont
  {Yamauchi}, \citenamefont {Sekiguchi}, \citenamefont {Chida}, \citenamefont
  {Arakawa}, \citenamefont {Nakamura}, \citenamefont {Kobayashi}, \citenamefont
  {Ono}, \citenamefont {Fujii},\ and\ \citenamefont
  {Sakano}}]{PhysRevLett.106.176601}%
  \BibitemOpen
  \bibfield  {author} {\bibinfo {author} {\bibfnamefont {Y.}~\bibnamefont
  {Yamauchi}}, \bibinfo {author} {\bibfnamefont {K.}~\bibnamefont {Sekiguchi}},
  \bibinfo {author} {\bibfnamefont {K.}~\bibnamefont {Chida}}, \bibinfo
  {author} {\bibfnamefont {T.}~\bibnamefont {Arakawa}}, \bibinfo {author}
  {\bibfnamefont {S.}~\bibnamefont {Nakamura}}, \bibinfo {author}
  {\bibfnamefont {K.}~\bibnamefont {Kobayashi}}, \bibinfo {author}
  {\bibfnamefont {T.}~\bibnamefont {Ono}}, \bibinfo {author} {\bibfnamefont
  {T.}~\bibnamefont {Fujii}}, and \bibinfo {author} {\bibfnamefont
  {R.}~\bibnamefont {Sakano}},\ }\href {\doibase
  10.1103/PhysRevLett.106.176601} {\bibfield  {journal} {\bibinfo  {journal}
  {Phys. Rev. Lett.}\ }\textbf {\bibinfo {volume} {106}},\ \bibinfo {pages}
  {176601} (\bibinfo {year} {2011})}\BibitemShut {NoStop}%
\bibitem [{\citenamefont {Ferrier}\ \emph {et~al.}(2016)\citenamefont
  {Ferrier}, \citenamefont {Arakawa}, \citenamefont {Hata}, \citenamefont
  {Fujiwara}, \citenamefont {Delagrange}, \citenamefont {Weil}, \citenamefont
  {Deblock}, \citenamefont {Sakano}, \citenamefont {Oguri},\ and\ \citenamefont
  {Kobayashi}}]{FerrierNatPhys2016}%
  \BibitemOpen
  \bibfield  {author} {\bibinfo {author} {\bibfnamefont {M.}~\bibnamefont
  {Ferrier}}, \bibinfo {author} {\bibfnamefont {T.}~\bibnamefont {Arakawa}},
  \bibinfo {author} {\bibfnamefont {T.}~\bibnamefont {Hata}}, \bibinfo {author}
  {\bibfnamefont {R.}~\bibnamefont {Fujiwara}}, \bibinfo {author}
  {\bibfnamefont {R.}~\bibnamefont {Delagrange}}, \bibinfo {author}
  {\bibfnamefont {R.}~\bibnamefont {Weil}}, \bibinfo {author} {\bibfnamefont
  {R.}~\bibnamefont {Deblock}}, \bibinfo {author} {\bibfnamefont
  {R.}~\bibnamefont {Sakano}}, \bibinfo {author} {\bibfnamefont
  {A.}~\bibnamefont {Oguri}}, and \bibinfo {author} {\bibfnamefont
  {K.}~\bibnamefont {Kobayashi}},\ }\href {\doibase 10.1038/nphys3556
  http://www.nature.com/nphys/journal/v12/n3/abs/nphys3556.html#supplementary-information}
  {\bibfield  {journal} {\bibinfo  {journal} {Nat. Phys.}\ }\textbf {\bibinfo
  {volume} {12}},\ \bibinfo {pages} {230} (\bibinfo {year} {2016})}\BibitemShut
  {NoStop}%
\bibitem [{\citenamefont {Liu}\ \emph {et~al.}(1998)\citenamefont {Liu},
  \citenamefont {Odom}, \citenamefont {Yamamoto},\ and\ \citenamefont
  {Tarucha}}]{liu}%
  \BibitemOpen
  \bibfield  {author} {\bibinfo {author} {\bibfnamefont {R.~C.}\ \bibnamefont
  {Liu}}, \bibinfo {author} {\bibfnamefont {B.}~\bibnamefont {Odom}}, \bibinfo
  {author} {\bibfnamefont {Y.}~\bibnamefont {Yamamoto}}, and \bibinfo
  {author} {\bibfnamefont {S.}~\bibnamefont {Tarucha}},\ }\href@noop {}
  {\bibfield  {journal} {\bibinfo  {journal} {Nature (London)}\ }\textbf {\bibinfo
  {volume} {391}},\ \bibinfo {pages} {263} (\bibinfo {year}
  {1998})}\BibitemShut {NoStop}%
\bibitem [{\citenamefont {Roche}\ \emph {et~al.}(2004)\citenamefont {Roche},
  \citenamefont {S{\'e}gala}, \citenamefont {Glattli}, \citenamefont
  {Nicholls}, \citenamefont {Pepper}, \citenamefont {Graham}, \citenamefont
  {Thomas},\ and\ \citenamefont {M.~Y.~Simmons}}]{roche}%
  \BibitemOpen
  \bibfield  {author} {\bibinfo {author} {\bibfnamefont {P.}~\bibnamefont
  {Roche}}, \bibinfo {author} {\bibfnamefont {J.}~\bibnamefont {S{\'e}gala}},
  \bibinfo {author} {\bibfnamefont {D.~C.}\ \bibnamefont {Glattli}}, \bibinfo
  {author} {\bibfnamefont {J.~T.}\ \bibnamefont {Nicholls}}, \bibinfo {author}
  {\bibfnamefont {M.}~\bibnamefont {Pepper}}, \bibinfo {author} {\bibfnamefont
  {A.~C.}\ \bibnamefont {Graham}}, \bibinfo {author} {\bibfnamefont {K.~J.}\
  \bibnamefont {Thomas}}, \bibinfo {author} {\bibfnamefont {M.~Y.~Simmons,}\, and \bibnamefont {D. A. Richie}},\ }\href@noop {} {\bibfield  {journal}
  {\bibinfo  {journal} {Phys. Rev. Lett.}\ }\textbf {\bibinfo {volume} {93}},\
  \bibinfo {pages} {116602} (\bibinfo {year} {2004})}\BibitemShut {NoStop}%
\bibitem [{\citenamefont {DiCarlo}\ \emph {et~al.}(2006)\citenamefont
  {DiCarlo}, \citenamefont {Zhang}, \citenamefont {McClure}, \citenamefont
  {Reilly},and \citenamefont {Marcus}}]{dicarlo}%
  \BibitemOpen
  \bibfield  {author} {\bibinfo {author} {\bibfnamefont {L.}~\bibnamefont
  {DiCarlo}}, \bibinfo {author} {\bibfnamefont {Y.}~\bibnamefont {Zhang}},
  \bibinfo {author} {\bibfnamefont {D.~T.}\ \bibnamefont {McClure}}, \bibinfo
  {author} {\bibfnamefont {D.~J.}\ \bibnamefont {Reilly}}, and \bibinfo
  {author} {\bibfnamefont {C.~M.}\ \bibnamefont {Marcus}},\ }\href@noop {}
  {\bibfield  {journal} {\bibinfo  {journal} {Phys. Rev. Lett.}\ }\textbf
  {\bibinfo {volume} {97}},\ \bibinfo {pages} {036810} (\bibinfo {year}
  {2006})}\BibitemShut {NoStop}%
\bibitem [{\citenamefont {Nakamura}\ \emph {et~al.}(2009)\citenamefont
  {Nakamura}, \citenamefont {Hashisaka}, \citenamefont {Yamauchi},
  \citenamefont {Kasai}, \citenamefont {Ono},\ and\ \citenamefont
  {Kobayashi}}]{nakamura}%
  \BibitemOpen
  \bibfield  {author} {\bibinfo {author} {\bibfnamefont {S.}~\bibnamefont
  {Nakamura}}, \bibinfo {author} {\bibfnamefont {M.}~\bibnamefont {Hashisaka}},
  \bibinfo {author} {\bibfnamefont {Y.}~\bibnamefont {Yamauchi}}, \bibinfo
  {author} {\bibfnamefont {S.}~\bibnamefont {Kasai}}, \bibinfo {author}
  {\bibfnamefont {T.}~\bibnamefont {Ono}}, and \bibinfo {author}
  {\bibfnamefont {K.}~\bibnamefont {Kobayashi}},\ }\href@noop {} {\bibfield
  {journal} {\bibinfo  {journal} {Phys. Rev. B}\ }\textbf {\bibinfo {volume}
  {79}},\ \bibinfo {pages} {201308} (\bibinfo {year} {2009})}\BibitemShut
  {NoStop}%
\bibitem [{\citenamefont {Kohda}\ \emph {et~al.}(2012)\citenamefont {Kohda},
  \citenamefont {Nakamura}, \citenamefont {Nishihara}, \citenamefont
  {Kobayashi}, \citenamefont {Ono}, \citenamefont {ichiro Ohe}, \citenamefont
  {Tokura}, \citenamefont {Mineno},\ and\ \citenamefont {Nitta}}]{kohda}%
  \BibitemOpen
  \bibfield  {author} {\bibinfo {author} {\bibfnamefont {M.}~\bibnamefont
  {Kohda}}, \bibinfo {author} {\bibfnamefont {S.}~\bibnamefont {Nakamura}},
  \bibinfo {author} {\bibfnamefont {Y.}~\bibnamefont {Nishihara}}, \bibinfo
  {author} {\bibfnamefont {K.}~\bibnamefont {Kobayashi}}, \bibinfo {author}
  {\bibfnamefont {T.}~\bibnamefont {Ono}}, \bibinfo {author} {\bibfnamefont
  {J.}~\bibnamefont {ichiro Ohe}}, \bibinfo {author} {\bibfnamefont
  {Y.}~\bibnamefont {Tokura}}, \bibinfo {author} {\bibfnamefont
  {T.}~\bibnamefont {Mineno}}, and \bibinfo {author} {\bibfnamefont
  {J.}~\bibnamefont {Nitta}},\ }\href@noop {} {\bibfield  {journal} {\bibinfo
  {journal} {Nat. Commn.}\ }\textbf {\bibinfo {volume} {3}},\ \bibinfo {pages}
  {1082} (\bibinfo {year} {2012})}\BibitemShut {NoStop}%
\bibitem [{\citenamefont {Kumar}\ \emph {et~al.}(1995)\citenamefont {Kumar},
  \citenamefont {Saminadayar}, \citenamefont {Glattli}, \citenamefont {Jin},\
  and\ \citenamefont {Etienne}}]{kumar}%
  \BibitemOpen
  \bibfield  {author} {\bibinfo {author} {\bibfnamefont {A.}~\bibnamefont
  {Kumar}}, \bibinfo {author} {\bibfnamefont {L.}~\bibnamefont {Saminadayar}},
  \bibinfo {author} {\bibfnamefont {D.~C.}\ \bibnamefont {Glattli}}, \bibinfo
  {author} {\bibfnamefont {Y.}~\bibnamefont {Jin}}, and \bibinfo {author}
  {\bibfnamefont {B.}~\bibnamefont {Etienne}},\ }\href@noop {} {\bibfield
  {journal} {\bibinfo  {journal} {Phys. Rev. Lett.}\ }\textbf {\bibinfo
  {volume} {76}},\ \bibinfo {pages} {2778} (\bibinfo {year}
  {1995})}\BibitemShut {NoStop}%
\bibitem [{\citenamefont {Reznikov}\ \emph {et~al.}(1995)\citenamefont
  {Reznikov}, \citenamefont {Heiblum}, \citenamefont {Shtrikman},\ and\
  \citenamefont {Mahalu}}]{reznikov}%
  \BibitemOpen
  \bibfield  {author} {\bibinfo {author} {\bibfnamefont {M.}~\bibnamefont
  {Reznikov}}, \bibinfo {author} {\bibfnamefont {M.}~\bibnamefont {Heiblum}},
  \bibinfo {author} {\bibfnamefont {H.}~\bibnamefont {Shtrikman}}, and
  \bibinfo {author} {\bibfnamefont {D.}~\bibnamefont {Mahalu}},\ }\href@noop {}
  {\bibfield  {journal} {\bibinfo  {journal} {Phys. Rev. Lett.}\ }\textbf
  {\bibinfo {volume} {75}},\ \bibinfo {pages} {3340} (\bibinfo {year}
  {1995})}\BibitemShut {NoStop}%
\bibitem [{\citenamefont {Hashisaka}\ \emph {et~al.}(2008)\citenamefont
  {Hashisaka}, \citenamefont {Yamauchi}, \citenamefont {Nakamura},
  \citenamefont {Kasai}, \citenamefont {Ono},\ and\ \citenamefont
  {Kobayashi}}]{hashisaka}%
  \BibitemOpen
  \bibfield  {author} {\bibinfo {author} {\bibfnamefont {M.}~\bibnamefont
  {Hashisaka}}, \bibinfo {author} {\bibfnamefont {Y.}~\bibnamefont {Yamauchi}},
  \bibinfo {author} {\bibfnamefont {S.}~\bibnamefont {Nakamura}}, \bibinfo
  {author} {\bibfnamefont {S.}~\bibnamefont {Kasai}}, \bibinfo {author}
  {\bibfnamefont {T.}~\bibnamefont {Ono}}, and \bibinfo {author}
  {\bibfnamefont {K.}~\bibnamefont {Kobayashi}},\ }\href@noop {} {\bibfield
  {journal} {\bibinfo  {journal} {Phys. Rev. B}\ }\textbf {\bibinfo {volume}
  {78}},\ \bibinfo {pages} {241303(R)} (\bibinfo {year} {2008})}\BibitemShut
  {NoStop}%
\bibitem [{\citenamefont {Nishihara}\ \emph {et~al.}(2012)\citenamefont
  {Nishihara}, \citenamefont {Nakamura}, \citenamefont {Kobayashi},
  \citenamefont {Ono}, \citenamefont {Kohda},\ and\ \citenamefont
  {Nitta}}]{nishihara}%
  \BibitemOpen
  \bibfield  {author} {\bibinfo {author} {\bibfnamefont {Y.}~\bibnamefont
  {Nishihara}}, \bibinfo {author} {\bibfnamefont {S.}~\bibnamefont {Nakamura}},
  \bibinfo {author} {\bibfnamefont {K.}~\bibnamefont {Kobayashi}}, \bibinfo
  {author} {\bibfnamefont {T.}~\bibnamefont {Ono}}, \bibinfo {author}
  {\bibfnamefont {M.}~\bibnamefont {Kohda}}, and \bibinfo {author}
  {\bibfnamefont {J.}~\bibnamefont {Nitta}},\ }\href@noop {} {\bibfield
  {journal} {\bibinfo  {journal} {Appl. Phys. Lett.}\ }\textbf {\bibinfo
  {volume} {100}},\ \bibinfo {pages} {203111} (\bibinfo {year}
  {2012})}\BibitemShut {NoStop}%
\bibitem [{\citenamefont {R{\"o}ssler}\ \emph {et~al.}(2011)\citenamefont
  {R{\"o}ssler}, \citenamefont {Baer}, \citenamefont {de~Wiljes}, \citenamefont
  {Ardelt}, \citenamefont {Ihn}, \citenamefont {Ensslin}, \citenamefont
  {Reichl},\ and\ \citenamefont {Wegscheider}}]{rossler}%
  \BibitemOpen
  \bibfield  {author} {\bibinfo {author} {\bibfnamefont {C.}~\bibnamefont
  {R{\"o}ssler}}, \bibinfo {author} {\bibfnamefont {S.}~\bibnamefont {Baer}},
  \bibinfo {author} {\bibfnamefont {E.}~\bibnamefont {de~Wiljes}}, \bibinfo
  {author} {\bibfnamefont {P.~L.}\ \bibnamefont {Ardelt}}, \bibinfo {author}
  {\bibfnamefont {T.}~\bibnamefont {Ihn}}, \bibinfo {author} {\bibfnamefont
  {K.}~\bibnamefont {Ensslin}}, \bibinfo {author} {\bibfnamefont
  {C.}~\bibnamefont {Reichl}}, and \bibinfo {author} {\bibfnamefont
  {W.}~\bibnamefont {Wegscheider}},\ }\href {\doibase
  10.1088/1367-2630/13/11/113006} {\bibfield  {journal} {\bibinfo  {journal}
  {New J. Phys.}\ }\textbf {\bibinfo {volume} {13}},\ \bibinfo {pages} {113006}
  (\bibinfo {year} {2011})}\BibitemShut {NoStop}%
\bibitem [{\citenamefont {R{\"o}ssler}\ \emph {et~al.}(2010)\citenamefont
  {R{\"o}ssler}, \citenamefont {Feil}, \citenamefont {Mensch}, \citenamefont
  {Ihn}, \citenamefont {Ensslin}, \citenamefont {Schuh},\ and\ \citenamefont
  {Wegscheider}}]{rossler2}%
  \BibitemOpen
  \bibfield  {author} {\bibinfo {author} {\bibfnamefont {C.}~\bibnamefont
  {R{\"o}ssler}}, \bibinfo {author} {\bibfnamefont {T.}~\bibnamefont {Feil}},
  \bibinfo {author} {\bibfnamefont {P.}~\bibnamefont {Mensch}}, \bibinfo
  {author} {\bibfnamefont {T.}~\bibnamefont {Ihn}}, \bibinfo {author}
  {\bibfnamefont {K.}~\bibnamefont {Ensslin}}, \bibinfo {author} {\bibfnamefont
  {D.}~\bibnamefont {Schuh}}, and \bibinfo {author} {\bibfnamefont
  {W.}~\bibnamefont {Wegscheider}},\ }\href
  {http://stacks.iop.org/1367-2630/12/i=4/a=043007} {\bibfield  {journal}
  {\bibinfo  {journal} {New J. Phys.}\ }\textbf {\bibinfo {volume} {12}},\
  \bibinfo {pages} {043007} (\bibinfo {year} {2010})}\BibitemShut {NoStop}%
\bibitem [{\citenamefont {Arakawa}\ \emph {et~al.}(2013)\citenamefont
  {Arakawa}, \citenamefont {Nishihara}, \citenamefont {Maeda}, \citenamefont
  {Norimoto},\ and\ \citenamefont {Kobayashi}}]{arakawa}%
  \BibitemOpen
  \bibfield  {author} {\bibinfo {author} {\bibfnamefont {T.}~\bibnamefont
  {Arakawa}}, \bibinfo {author} {\bibfnamefont {Y.}~\bibnamefont {Nishihara}},
  \bibinfo {author} {\bibfnamefont {M.}~\bibnamefont {Maeda}}, \bibinfo
  {author} {\bibfnamefont {S.}~\bibnamefont {Norimoto}}, and \bibinfo
  {author} {\bibfnamefont {K.}~\bibnamefont {Kobayashi}},\ }\href@noop {}
  {\bibfield  {journal} {\bibinfo  {journal} {Appl. Phys. Lett.}\ }\textbf
  {\bibinfo {volume} {103}},\ \bibinfo {pages} {172104} (\bibinfo {year}
  {2013})}\BibitemShut {NoStop}%
\bibitem [{\citenamefont {Steinbach}\ \emph {et~al.}(1996)\citenamefont
  {Steinbach}, \citenamefont {Martinis},\ and\ \citenamefont
  {Devoret}}]{steinbach}%
  \BibitemOpen
  \bibfield  {author} {\bibinfo {author} {\bibfnamefont {A.~H.}\ \bibnamefont
  {Steinbach}}, \bibinfo {author} {\bibfnamefont {J.~M.}\ \bibnamefont
  {Martinis}}, and \bibinfo {author} {\bibfnamefont {M.~H.}\ \bibnamefont
  {Devoret}},\ }\href@noop {} {\bibfield  {journal} {\bibinfo  {journal} {Phys.
  Rev. Lett.}\ }\textbf {\bibinfo {volume} {76}},\ \bibinfo {pages} {3806}
  (\bibinfo {year} {1996})}\BibitemShut {NoStop}%
\bibitem [{\citenamefont {Dzurak}\ \emph {et~al.}(1992)\citenamefont {Dzurak},
  \citenamefont {Ford}, \citenamefont {Kelly}, \citenamefont {Pepper},
  \citenamefont {Frost}, \citenamefont {Ritchie}, \citenamefont {Jones},
  \citenamefont {Ahmed},\ and\ \citenamefont {Hasko}}]{dzurak}%
  \BibitemOpen
  \bibfield  {author} {\bibinfo {author} {\bibfnamefont {A.~S.}\ \bibnamefont
  {Dzurak}}, \bibinfo {author} {\bibfnamefont {C.~J.~B.}\ \bibnamefont {Ford}},
  \bibinfo {author} {\bibfnamefont {M.~J.}\ \bibnamefont {Kelly}}, \bibinfo
  {author} {\bibfnamefont {M.}~\bibnamefont {Pepper}}, \bibinfo {author}
  {\bibfnamefont {J.~E.~F.}\ \bibnamefont {Frost}}, \bibinfo {author}
  {\bibfnamefont {D.~A.}\ \bibnamefont {Ritchie}}, \bibinfo {author}
  {\bibfnamefont {G.~A.~C.}\ \bibnamefont {Jones}}, \bibinfo {author}
  {\bibfnamefont {H.}~\bibnamefont {Ahmed}}, and \bibinfo {author}
  {\bibfnamefont {D.~G.}\ \bibnamefont {Hasko}},\ }\href@noop {} {\bibfield
  {journal} {\bibinfo  {journal} {Phys. Rev. B}\ }\textbf {\bibinfo {volume}
  {45}},\ \bibinfo {pages} {6309(R)} (\bibinfo {year} {1992})}\BibitemShut
  {NoStop}%
\bibitem [{\citenamefont {Hirayama}\ \emph {et~al.}(2013)\citenamefont
  {Hirayama}, \citenamefont {Endo}, \citenamefont {Fujita}, \citenamefont
  {Hasegawa}, \citenamefont {Hatano}, \citenamefont {Nakamura}, \citenamefont
  {Shirasaki},\ and\ \citenamefont {Yonemitsu}}]{HirayamaJLTP2013}%
  \BibitemOpen
  \bibfield  {author} {\bibinfo {author} {\bibfnamefont {N.}~\bibnamefont
  {Hirayama}}, \bibinfo {author} {\bibfnamefont {A.}~\bibnamefont {Endo}},
  \bibinfo {author} {\bibfnamefont {K.}~\bibnamefont {Fujita}}, \bibinfo
  {author} {\bibfnamefont {Y.}~\bibnamefont {Hasegawa}}, \bibinfo {author}
  {\bibfnamefont {N.}~\bibnamefont {Hatano}}, \bibinfo {author} {\bibfnamefont
  {H.}~\bibnamefont {Nakamura}}, \bibinfo {author} {\bibfnamefont
  {R.}~\bibnamefont {Shirasaki}}, and \bibinfo {author} {\bibfnamefont
  {K.}~\bibnamefont {Yonemitsu}},\ }\href {\doibase 10.1007/s10909-012-0852-8}
  {\bibfield  {journal} {\bibinfo  {journal} {J. Low Temp. Phys.}\ }\textbf
  {\bibinfo {volume} {172}},\ \bibinfo {pages} {132} (\bibinfo {year}
  {2013})}\BibitemShut {NoStop}%
\end{thebibliography}

\providecommand{\noopsort}[1]{}\providecommand{\singleletter}[1]{#1}%

\end{document}